\documentclass[12pt]{article}
\usepackage{latexsym}
\usepackage{amsmath}
\usepackage{amssymb}
\usepackage{epsfig}
\usepackage{epic}

\numberwithin{equation}{subsection}
\numberwithin{figure}{subsection}

\begin{document}

\def\k{\kappa}
\def\half{\fraction{1}{2}}
\def\fraction#1#2{ { \scriptstyle \frac{#1}{#2} }}
\def\der#1{\frac{\partial}{\partial #1}}
\def\d{\partial}
\def\p2{\fraction{\pi}{2}}
\def\s{\sigma}
\def\Id{|I\rangle}
\def\Tr{\mathrm{Tr}}
\def\z{\xi}
\def\eq#1{eq.(\ref{eq:#1})}
\def\Eq#1{Eq.(\ref{eq:#1})}

\begin{titlepage}
\rightline{\today}

\begin{center}
\vskip 1.0cm
{\Large \bf{Split String Formalism and the Closed String Vacuum} }
\\
\vskip 1.0cm

{\large Theodore Erler}
\vskip 1.0cm
{\it {Harish-Chandra Research Institute} \\ 
{Chhatnag Road, Jhunsi, Allahabad 211019, India}}
\\ E-mail:terler@mri.ernet.in \\

\vskip 1.0cm
{\bf Abstract}
\end{center}

\noindent
The split string formalism offers a simple template upon which we 
can build many generalizations of Schnabl's analytic solution of open 
string field theory. In this paper we consider two such generalizations: one 
which replaces the wedge state by an arbitrary function of wedge states, and 
another which generalizes the solution to conformal frames other than the 
sliver.

\medskip

\end{titlepage}

\newpage

\baselineskip=18pt

\tableofcontents

\section{Introduction}

In a remarkable paper\cite{Schnabl} Schnabl constructed, for the first time, 
a well-behaved analytic solution to the equations of motion of open bosonic 
string field theory (OSFT)\footnote{For a nice review, see ref.\cite{Review}.}.
The solution represents the most basic and 
important field configuration for the open string: the vacuum where the D-brane
has decayed and there are no open strings left. By now it has been confirmed
numerically and analytically that the potential energy of the vacuum matches
the D-brane tension\cite{Schnabl,Okawa,Fuchs} and that the vacuum supports 
no open strings\cite{Ellwood}, in accordance with Sen's conjectures\cite{Sen}.

Given the complexity of the solution in its various forms, it is instructive
to strip it down to its bare bones in the hope generalizing its structure and
codifying essential lessons in the search for other solutions. As noticed by
Okawa\cite{Okawa} perhaps the simplest expression of Schnabl's solution
comes in the split string formalism\cite{Gross-Taylor,Moyal}. There, the 
solution can be expressed algebraically in terms of ``matrix products'' of 
three string fields $K,B$ and $c$. These three fields are postulated to 
satisfy six simple identities \eq{SSid}. The expression,
\begin{equation}\Psi = F(K)c\frac{KB}{1-F(K)^2}c F(K)\label{eq:Sch_intro}
\end{equation}
is then guaranteed to satisfy the equations of motion for any string field $F$ 
which depends only on $K$. For Schnabl's solution, $F(K)$ is a wedge state,
specifically the ``square root'' of the $SL(2,\mathbb{R})$ invariant vacuum.
 
In this paper we investigate this basic structure in the search for 
generalizations of Schnabl's solution. We consider two types of 
generalization: the first generalizes the choice of the field $F(K)$, and the
second searches for new realizations $K,B,c$ such that the simple identities 
which make \eq{Sch_intro} work are still satisfied. This second 
generalization, we will see, is essentially equivalent to the choice of a 
projector and its associated conformal frame. Schnabl's solution is based 
on the sliver projector, the state obtained by repeated multiplication of the
$SL(2,\mathbb{R})$ vacuum with itself. Aside from some general comments, our 
discussion of generic $F(K)$ will be limited; we will analyze the simplest 
example where $F$ is allowed to be an arbitrary wedge state. The completely 
general case will be considered in a companion paper\cite{Erler} (henceforth, 
(II)).

This paper is organized as follows. In section \ref{sec:SSF} we develop the 
basic framework and explain how Schnabl's solution maps between the split
string formalism and the conformal field theory representation. In section
\ref{sec:F} we slightly generalize $F(K)$, allowing it to be an arbitrary 
power of the $SL(2,\mathbb{R})$ vacuum; we show that these solutions are 
related by a simple midpoint-preserving reparameterization symmetry. In 
section \ref{sec:CFT}, we generalize Schnabl's solution to other projector 
conformal frames. The major challenge 
here is to find a workable conformal field theory representation of these 
solutions. Our strategy is to define a new conformal frame, the ``strip 
frame,'' where the equivalent of wedge states are described as infinite
strips in the complex plane. Manipulation of the solution and calculation
of the energy then proceeds in exact analogy with Schnabl's solution. We
illustrate this for the butterfly projector and some of its multi-winged 
cousins. We end with some conclusions.

While our work was nearing completion, the paper ref.\cite{RZO} appeared 
which has significant overlap with some of our results. We hope however that 
this paper offers a useful perspective on these types of generalization.

\section{Split String Formalism}
\label{sec:SSF}

The split string formalism (SSF) is a formal approach to computations 
in OSFT which represents the star product as a ``matrix product'' of half 
string functionals\cite{Gross-Taylor,Moyal}. The basic idea is quite 
intuitive, though the concrete implementation can become quite technical.
For our purposes, however, all we need is the basic philosophy: that the action
of an operator on the string field $\Psi$ can represented as star 
multiplication of $\Psi$ with an appropriate state. In this way, all 
computations in OSFT reduce to computations of the star product 
(``matrix product'') and the BPZ inner product (``trace'').

The starting point for the SSF is the overlap condition, which states that 
the vertex identifies the left and right halves of the string for 
neighboring functionals, defining a kind of ``matrix product.'' In particular,
if $\langle \Psi_1,\Psi_2*...\rangle_N$ is the $N$-string vertex, the string 
position coordinate $x(\s)$ satisfies,
\begin{eqnarray}\langle \Psi_1,x(\s)\Psi_2*...\rangle_N &=& 
\langle x(\pi-\s)\Psi_1,\Psi_2*...\rangle_N\ \ \ \s\in[0,\p2]\nonumber\\
&\ & x(\s) = X(z,\bar{z})|_{z=e^{i\s}}\nonumber
\end{eqnarray}
and so on cyclically. This can be generalized to an arbitrary local operator 
$A(z)$. Let
$$\phi=e^{i\s}\ \ \ \s\in[-\p2,\p2] $$
so that $A(\phi)$ acts on the left half of the string at $\s$. Then the 
overlap condition states\footnote{If $A$ is Grassmann odd we must
multiply by the appropriate sign from anticommutation.},
\begin{equation}\langle \Psi_1,A(\phi)\Psi_2*...\rangle_N = 
\langle A^*(\phi)\Psi_1,\Psi_2*...\rangle_N
\label{eq:ov_cond}\end{equation}
where $A^*$ is the BPZ dual of $A$. If $A$ is a dimension $h$ 
primary, then
\begin{equation}A^*(\phi) = 
\phi^{-2h}\bar{\phi}^{-2\bar{h}}A\left(-\frac{1}{\phi}\right)\end{equation}
Another useful object is the identity string field $\Id$, which is postulated
to satisfy
\begin{equation}\langle I,\Psi_1*\Psi_2*...\rangle_{N+1} = 
\langle \Psi_1,\Psi_2*...\rangle_N\end{equation}
The existence of $\Id$ immediately gives a prescription for implementing
the action of operators on the star algebra in terms of the star algebra
itself. For example we can write, 
\begin{equation}A(\phi)\Psi = \mathcal{A}*\Psi
\end{equation} 
where the field $\mathcal{A}$ is defined,
\begin{equation}
\mathcal{A} = A(\phi)\Id
\end{equation}
The proof is,
\begin{eqnarray}\langle \chi, A(\phi)\Psi\rangle_2 &=& 
\langle A^*(\phi)\chi, \Psi\rangle_2 = \langle A^*(\phi)\chi,
I*\Psi\rangle_3 = \langle \chi, (A(\phi)I)*\Psi\rangle_3\nonumber\\
&=& \langle \chi,\mathcal{A}*\Psi\rangle_3\label{eq:op_star}
\end{eqnarray}
A similar argument shows that for operators acting on the right half
of the string,
\begin{equation}A^*(\phi)\Psi = (-1)^{A\Psi}\Psi*\mathcal{A}
\end{equation} Also,
\begin{equation}A^*(\phi)\Id = A(\phi)\Id
\label{eq:Id_ov}\end{equation}
Note that the above manipulations become a little delicate when $A$ acts on 
the midpoint $\phi=\pm i$. Unless $A$ has dimension zero, 
evaluating $A(i)$ in the vertex creates vanishing or divergent factors which 
can yield some surprises. 

Let us make some remarks about conventions. In our definition of the 
vertex \eq{ov_cond} the right string half of the previous functional 
is identified with the left string half of the subsequent functional. 
Sometimes the opposite convention is used, for example in ref.\cite{Okawa}; 
these conventions are related by a twist. Also, the left half of the string
$\s\in[0,\p2]$ is mapped to the {\it right} half semi-circle $e^{i\s}$ in the
complex plane. To avoid confusion, we will always refer to the image of the
left half of the string in the complex plane as the {\it positive} side, 
and the image of the right as the {\it negative} side.

The only other thing we need from the SSF is notation. When writing the star 
product we drop the star, and when calculating the BPZ inner product we replace
the bracket with a trace. Thus,
\begin{equation}\Psi*\Phi = \Psi\Phi\ \ \ \ \langle \Psi,\Phi\rangle = 
\Tr(\Psi\Phi)\end{equation}
The identity string field is denoted by $1$. We may then manipulate string 
fields like ordinary matrices, taking care of Grassmann parity. 
For example,
\begin{equation}\Tr(\Psi\Phi) = (-1)^{\Psi\Phi}
\Tr(\Phi\Psi)\end{equation}
The physical string field is Grassmann odd.

\subsection{Solution in the SSF}

As noticed by Okawa\cite{Okawa}, we can construct a ``split string'' solution 
to the field equations given any three string fields,
\begin{eqnarray}K &=& \mathrm{Grassmann\ even,\ gh}\#= 0\nonumber\\ 
B &=& \mathrm{Grassmann\ odd,\ gh}\#= -1\nonumber\\
c &=& \mathrm{Grassmann\ odd,\ gh}\#= 1\end{eqnarray}
which satisfy the following algebraic properties:
\begin{eqnarray}&\ &Bc + cB =1\ \ \ \ \ \ \ \ KB - BK = 0 \ \ \ \ \ \ 
B^2 = c^2 = 0\nonumber\\ &\ &\ \ \ \ \ \ \ 
dK = 0 \ \ \ \ \ \ \ \ \ \ dB = K \ \ \ \ 
\ \ \ \ \ dc = cKc\label{eq:SSid}\end{eqnarray}
where $d=Q_B$ is the BRST operator\footnote{In the split string formalism we 
might like to express $d$ as an inner derivation 
$d\Psi = D\Psi -(-1)^{\epsilon(\Psi)}\Psi D$ for some string field $D$. 
Formally this is possible\cite{Horowitz}, though there can be subtleties 
because $d$ acts on the midpoint. We will not find it necessary or useful to 
express $d$ as an inner derivation.}. These relations imply that $K,B,c$ 
freely generate a 
subalgebra of the star algebra which is closed under the action of the BRST
operator. It is then simple to show\cite{Okawa} that the state,
\begin{equation}\Psi = F(K)c\frac{KB}{1-F^2(K)}c F(K)\label{eq:SSsol}
\end{equation}
formally satisfies,\begin{equation}d\Psi + \Psi^2 =0\end{equation}
for any field $F(K)$ which can be written purely 
in terms of star products of $K$ alone. For the appropriate choice of 
$K,B,c$ and $F$, this is in fact Schnabl's solution\footnote{It seems possible
that eq.\ref{eq:SSsol} describes all acceptable solutions within the 
subalgebra generated by $K,B,c$. We have found another solution 
$\Psi = Fc\frac{K}{F}$ but it seems singular and does not satisfy the reality 
condition. If other solutions exist it could be of great interest.}.

\begin{figure}[top]

\begin{center}
\resizebox{2.3in}{2.0in}{\includegraphics{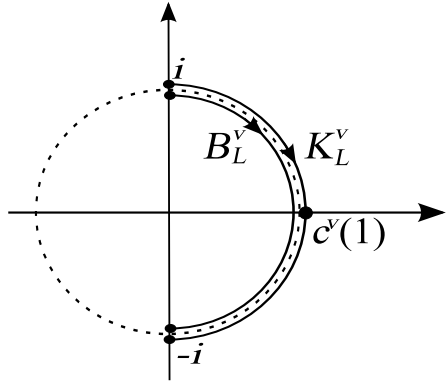}}
\end{center}
\caption{Operators defining the fields $K,B,c$.\label{fig:KBc}}

 .\hrulefill
\end{figure}

Let us see how the identities \eq{SSid} can be explicitly realized. 
We will suppose $K,B,c$ take the form
\begin{eqnarray}
K &=& K_L^v\Id\nonumber\\
B &=& B_L^v\Id\nonumber\\
c &=& c^v(1)\Id\label{eq:KBc}
\end{eqnarray}
where,
\begin{equation}K_L^v = \int_L \frac{d\z}{2\pi i} v(\z)T(\z)\ \ \ \ \ 
B_L^v = \int_L \frac{d\z}{2\pi i} v(\z)b(\z)\ \ \ 
\ \ c^v(1) = -\frac{1}{v(1)}c(1)\label{eq:KBc_op}
\end{equation} and $v(\z)$ is an arbitrary holomorphic vector field, subject 
to certain conditions which we will explain in a moment. The contour $L$ 
is taken over the ``left half'' of the string, that is, along the positive 
semi-circle connecting $\z = +i$ to $\z =-i$ (see \ref{fig:KBc}). We will use
$R$ to denote the contour from $i$ to $-i$ on the negative semi-circle. From 
these definitions and \eq{op_star} it is straightforward to verify 
\eq{SSid}. For example,
\begin{equation}Bc+cB = ( B_L^vc^v(1)+ c^v(1)B_L^v )\Id = 1\end{equation}
and
\begin{equation}dB = (Q_B B_L^v +B_L^vQ_B)\Id = K_L^v\Id = K\end{equation}
where we used the fact that the identity is BRST closed.

The vector field $v(\z)$ is subject to two conditions: reality and regularity.
The reality condition follows from the requirement that $K,B,c$ be real string
fields, which in split string language means they are ``self-adjoint'' 
matrices, $K^\dag=K$ etc. If (and only if) this condition holds, the 
solution $\Psi$ is real. For a state of the form $\mathcal{O}\Id$, the 
reality condition reads,
$$\langle \mathcal{O}I,\chi\rangle = \langle I|\mathcal{O}^\dag
|\chi\rangle $$
Using \eq{Id_ov} and the fact that $\Id$ is real, 
we can write the left hand side, 
\begin{equation}\langle \mathcal{O}^*I,\chi\rangle = 
\langle I,\mathcal{O}\chi\rangle = \langle I|\mathcal{O}|\chi\rangle
\end{equation}
Thus $\mathcal{O}$ must be a hermitian operator. Imposing this on 
\eq{KBc_op} requires,
\begin{equation}\overline{v(\z)} = \bar{\z}^2 v\left(\frac{1}{\bar{\z}}\right)
\label{eq:v_real}\end{equation} 
Note that in our definition of $c$ we took the ghost 
insertion to be precisely on the real axis. Actually, if the insertion is taken
off the real axis the identities \eq{SSid} are still satisfied, but 
then $c$ and the resulting solution would not be real.

The second condition on $v(\z)$ is a regularity condition: $v$ must vanish
at the midpoint,
\begin{equation} v(\pm i)=0 \label{eq:v_reg}\end{equation}
As mentioned earlier, when representing operators as string fields in the SSF,
it is best to avoid operators which act on the midpoint. Furthermore, as 
explained in ref.\cite{RZ}, left/right decompositions of energy momentum
charges $\oint vT $ are anomalous unless $v(\pm i)=0$. Probably solutions 
which fail to satisfy \eq{v_reg} are undefined.

Following the philosophy of refs.\cite{Schnabl,RZ} it is useful to think of
the operator $K_L^v$ as arising from the energy momentum zero mode in a 
nonstandard conformal frame. Let $z=f(\z)$ define the local coordinate for 
a surface state represented on the upper half plane. The energy momentum
zero mode in this conformal frame is,
\begin{equation}\mathcal{L}_0 = \oint \frac{dz}{2\pi i}zT(z) = 
\oint\frac{d\z}{2\pi i}\frac{f(\z)}{f'(\z)}T(\z)
\end{equation}
Its BPZ conjugate is,
\begin{equation}\mathcal{L}_0^* = \oint \frac{d\z}{2\pi i} 
\left(-\z^2\frac{f(-1/\z)}{f'(-1/\z)}\right)T(\z)
\end{equation} Up to a proportionality we may identify,
\begin{equation}K_L^v = (\mathcal{L}_0+\mathcal{L}_0^*)_L
\label{eq:K_Lid}\end{equation}
where the subscript $L$ denotes integrating the corresponding current over 
the contour $L$ pictured in fig.(\ref{fig:KBc}\footnote{Note that 
$K_L^v=-sL^+_L$ in the notation of ref.\cite{RZ}. The sign comes from our 
definition of ``left,'' which is chosen so that $K_L^v$ creates, rather than 
destroys, infinitesimal strips in the sliver coordinate frame.}). Thus,
\begin{equation}v(\z) = \frac{f(\z)}{f'(\z)}-\z^2\frac{f(-1/\z)}{f'(-1/\z)}
\end{equation} Notice that for real, twist invariant conformal frames,
$$\overline{f(\z)} = f(\bar{\z})\ \ \ f(-\z) = -f(\z)$$
the reality condition \eq{v_real} is automatically satisfied. Moreover,
the regularity condition \eq{v_reg} follows if
\begin{equation}f(\pm i) =\infty\end{equation}
that is, if $f(\z)$ is the conformal frame for a 
{\it projector}\cite{Projectors}. Thus, the importance of projectors in
Schnabl's solution is ultimately related to the fact that it can be 
expressed using the split string formalism.

In summary, we have a generic set of solutions to the string field equation,
characterized by an arbitrary field $F(K)$ and choice of projector with 
conformal frame $f(\z)$. In the rest of this paper we study the implications of
these solutions. 

\subsection{Schnabl's solution}

Though it may be possible to realize the solution $\Psi$ in terms of a matrix
or Moyal product of open string functionals, our strategy will be to map the
solutions to the conformal field theory representation where it is clearer 
how to define $\Psi$ in terms of correlators.

\begin{figure}[top]
\begin{center}
\resizebox{3in}{2in}{\includegraphics{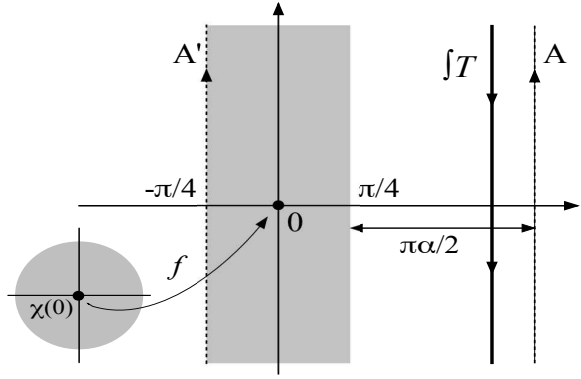}}
\end{center}
\caption{\label{fig:wedge}
Correlator defining the derivative of the wedge state 
$\Omega^\alpha$. The dashed vertical lines A,A' are cut and sewn together 
to form a cylinder.}
\end{figure}

In this capacity, let us review Schnabl's solution in the form 
\eq{SSsol} and its relation to the conformal field theory 
representation. Schnabl's solution corresponds to the choice,
\begin{eqnarray}f(\z) &=& \tan^{-1}\z \label{eq:sliver_cf}\\
F(K) &=& \exp(\half K) = \Omega^{1/2}\label{eq:SS_Sch_sol}\end{eqnarray}
The first equation says the solution is formulated in the sliver conformal 
frame\cite{Schnabl,Projectors}. The second equation says that $F$ is the 
square root of the $SL(2,\mathbb{R})$ vacuum $\Omega$. We also have,
\begin{eqnarray}v(\z) &=& \frac{\pi}{2}(1+\z^2)\nonumber\\
K_L^v &=& \frac{\pi}{2}(L_1 +L_{-1})_L\ \ \ B_L^v = 
\frac{\pi}{2}(b_1+b_{-1})_L\ \ \ 
c^v(1)=-\frac{1}{\pi}c(1)\label{eq:Sch_KBc}\end{eqnarray}
Let us recall why $F$ in the form \eq{SS_Sch_sol} is the square root
of the vacuum, because it's important. A wedge state $\Omega^\alpha$ can
be represented as a correlation function on the cylinder 
$C_{\frac{\pi}{2}(\alpha+1)}$. This cylinder can be represented as a strip
$-\frac{\pi}{4}<\Re(z)<\frac{\pi\alpha}{2}+\frac{\pi}{4}$ with its ends 
identified. Specifically,
\begin{equation}\langle \Omega^\alpha,\chi\rangle = \langle f\circ\chi(0)
\rangle_{C_{\frac{\pi(\alpha+1)}{2}}}\label{eq:wedge}\end{equation}
with $\chi(0)$ the vertex operator of the state $\chi$ and $f$ as in 
\eq{sliver_cf}. The derivative of $\Omega^{\alpha}$ with respect to 
$\alpha$ is the proportional to the variation of the correlator with respect
to the cylinder's circumference. Alternatively, this variation can be computed 
by inserting a ``Hamiltonian'' into the correlator which creates an 
infinitesimal strip of worldsheet parallel to the imaginary axis, as shown 
in figure \ref{fig:wedge}. Thus,
\begin{equation}\left\langle \frac{d}{d\alpha}\Omega^\alpha,\chi\right\rangle 
= 
\frac{\pi}{2}\frac{d}{d(\frac{\pi(\alpha+1)}{2})}\langle f\circ\chi(0)
\rangle_{C_{\frac{\pi(\alpha+1)}{2}}} = \frac{\pi}{2}
\left\langle f\circ\chi(0)\int_{i\infty}^{-i\infty}\frac{dz}{2\pi i}T(z)
\right\rangle_{C_{\frac{\pi(\alpha+1)}{2}}}\end{equation}
We now absorb the $\int T$ into the local coordinate on the {\it negative} 
side of the puncture, and map back to the unit disk. The result is,
\begin{equation}\left\langle \frac{d}{d\alpha}\Omega^\alpha,\chi\right\rangle 
= \left\langle f\circ\left(\chi(0)\int_R\frac{d\z}{2\pi i}\frac{\pi}{2}(1+\z^2)
T(\z)\right)
\right\rangle_{C_{\frac{\pi(\alpha+1)}{2}}}=\langle \Omega^{\alpha},
K_R^v\chi\rangle\end{equation} Noting $K_R^v{}^* = K_L^v$, this gives,
\begin{equation}\frac{d}{d\alpha}\Omega^\alpha = K_L^v\Omega^\alpha
\end{equation} Integrating and using \eq{op_star}
\begin{equation}\Omega^\alpha = \exp(\alpha K_L^v)\Id = \exp(\alpha K)
\end{equation}
implying \eq{SS_Sch_sol}\footnote{While the relation 
$\exp(\alpha K_L^v)\Id = \exp(\alpha K)$ appears formal, we have checked
that it can be given a precise meaning in the Moyal representation 
of the star product, at least in the matter sector. Ghosts would be trickier to
handle because one has to treat zero modes carefully and cancel off
divergent determinants. Perhaps the recent results of ref.\cite{regulator} 
would help in this regard.}.

\begin{figure}[top]
\begin{center}
\resizebox{3.6in}{1.9in}{\includegraphics{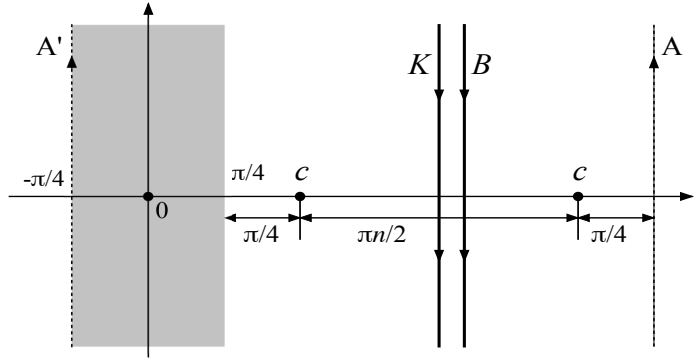}}
\end{center}
\caption{\label{fig:psi_cor}
Conformal field theory representation of the state 
$\Omega^\alpha cKB\Omega^\beta c\Omega^\gamma$.}
\end{figure}

To calculate the D-brane energy using Schnabl's solution, it is necessary to 
regulate the form \eq{SSsol} in a particular way\cite{Schnabl}.
Specifically, the solution is rewritten
\begin{equation}\Psi = \lim_{N\to\infty}\left(\sum_{n=0}^N \psi_n'- 
\psi_N
\right)\label{eq:SS_sol_reg}\end{equation}
where,
\begin{equation}\psi_n' = FcBKF^{2n}cF\ \ \ \ \ 
\psi_n = \frac{1}{\gamma}FcBF^{2n}cF\label{eq:psi_ns}
\end{equation}
and $\gamma$ is a constant associated with the definition of $F$, as we will 
explain in section \ref{sec:F}\footnote{For Schnabl's solution $\gamma=1$, 
as shown in \cite{Schnabl,Okawa,Fuchs}.}. This amounts to a 
truncated Taylor expansion of our 
previous formula, modulo the mysterious $\psi_N$ piece which vanishes when 
contracted with Fock space states\cite{Schnabl}. In the following we will
{\it assume} that this regulator is correct, regardless of the choice of 
$F$ or conformal frame\footnote{Note that for general $F$, 
$\psi_n'\neq\frac{d}{dn}\psi_n$.}. \Eq{SS_sol_reg} is also convenient 
for translating to CFT language, since when $F=\Omega^{1/2}$ the $\psi_n'$s 
can be computed as correlators on the cylinder with particular 
insertions\cite{Okawa}. To get the insertions and their normalizations 
correctly, it is useful to note that for an primary operator $A(\phi)$ acting 
on the left side of the string,
\begin{equation}\langle A(\phi)\Omega^\alpha,\chi\rangle = 
\frac{1}{(1+\phi^2)^{h}}\left\langle 
A\left(\frac{\pi(\alpha+\frac{1}{2})}{2}+iy\right)
f\circ\chi(0)\right\rangle_{C_{\frac{\pi(\alpha+1)}{2}}}\ \ \ 
\end{equation} 
where,
\begin{equation}y = \frac{1}{4}\tanh^{-1}\sin\sigma\end{equation}
In particular, the operator is mapped to the positive boundary of the strip 
defining the wedge state, $\Re(z)=\frac{\pi(\alpha+\frac{1}{2})}{2}$. The
string endpoint $\phi=1$ ($\sigma=0$) is mapped to the real axis, and the
midpoint $\phi=i$ ($\sigma=\frac{\pi}{2}$) is mapped to infinity, as expected.
Using this mapping, the definition of $K,B$ and $c$, and the CFT gluing 
prescriptions for the star product, one can rewrite the $\psi_n'$s as 
correlators on the cylinder: 
\begin{eqnarray} \langle\psi_n',\chi\rangle &=&
\left\langle c\left(\frac{\pi(n+1)}{2}\right)KB
c\left(\frac{\pi}{2}\right)f\circ\chi(0)
\right\rangle_{C_{\frac{\pi(n+2)}{2}}}
\label{eq:SSF_cylinder}\end{eqnarray}
where we have defined the contour insertions,
\begin{equation}K = \int_{i\infty}^{-i\infty}\frac{dz}{2\pi i}T(z)\ \ \ \ 
B = \int_{i\infty}^{-i\infty}\frac{dz}{2\pi i}b(z)
\label{eq:KB_cont}\end{equation} 
hopefully not to be confused the fields $K,B$ introduced 
earlier (see figure \ref{fig:psi_cor}). This is the well-known result of 
ref.\cite{Okawa}\footnote{Actually, perhaps the simplest formulation is to 
define the fields $K,B,c$ as cylinder correlators from the start. For example,
$$\langle c,\chi\rangle = - 
\left\langle c\left(\frac{\pi}{2}\right)f\circ\chi(0)\right
\rangle_{C_{\frac{\pi}{2}}}$$ Translating between CFT and the split string 
formalism then becomes trivial. However,
generalizing this to arbitrary projector frames requires new ingredients,
to be explained in section \ref{sec:CFT}.} 

\subsection{Energy: general considerations}
\label{subsec:Energy}

Ultimately we are interested in calculating the energy and giving a physical 
interpretation of these new solutions. Thus, it's worthwhile seeing what can 
be said about the energy before we specify $F,f$ and try to evaluate 
correlators. 

Assuming the equations of motion, the action evaluated on the (regulated) 
solution is,
\begin{eqnarray}E &=& -\frac{1}{6}\langle \Psi,Q_B\Psi\rangle\nonumber\\
&=& -\frac{1}{6}\lim_{N\to\infty}
\left[\sum_{m,n=0}^N \langle \psi_m',Q_B\psi'_n\rangle
-2\sum_{m=0}^N\langle\psi'_m,Q_B\psi_N\rangle 
+\langle \psi_N,Q_B\psi_N\rangle\right]\end{eqnarray}
A crucial step in evaluating this comes from the realization that,
for Schnabl's $\psi_n'$s, the following ``diagonal'' sum vanishes,
\begin{equation}\sum_{k=0}^n\langle\psi'_{n-k},Q_B\psi_k'\rangle =0 
\label{eq:diag_sum}\end{equation} 
Because of this, the action only receives contributions
from $\langle \psi_m,Q_B\psi_n\rangle$ for large $n+m$, where the correlators
simplify\cite{Schnabl,Okawa,Fuchs}. Furthermore, for the pure gauge 
solutions of \cite{Schnabl}, the large $n+m$ terms manifestly don't 
contribute, so \eq{diag_sum} implies their energies vanish. 

In ref.\cite{Schnabl}, \eq{diag_sum} was demonstrated by explicit 
evaluation of $\langle \psi'_m,Q_B\psi'_n\rangle$. As we will argue, this 
sum actually vanishes in general, regardless of the choice of $F$ and 
conformal frame. This could be demonstrated directly with the proper 
manipulations invoking the identities \eq{SSid} and OSFT axioms, but we found 
it difficult to construct a simple proof in this manner.
Instead, we utilize an observation of Okawa\cite{Okawa} that Schnabl's solution
can be recast in a form which is (naively) pure gauge. Define the state,
\begin{equation}\Phi = FBcF \end{equation}
A little computation with \eq{SSid} shows that,
\begin{equation}(d\Phi)\Phi^n = FcKBF^{2n}cF = \psi_n'\label{eq:Ok_Phi}
\end{equation}
Let us define a one parameter family of solutions,
\begin{equation}\Psi_\lambda = \lambda Fc\frac{KB}{1-\lambda F^2}cF = 
\sum_{n=0}^\infty \lambda^{n+1}\psi_n'
\label{eq:SSgauge_sol}\end{equation}
The real parameter $\lambda$ could be absorbed into the definition of $F$, but
let us keep it there for convenience. Plugging in \eq{Ok_Phi} gives,
\begin{equation}\Psi_\lambda = \lambda(d\Phi)\frac{1}{1-\lambda\Phi}
=(1-\lambda\Phi)d\left(\frac{1}{1-\lambda\Phi}\right)\end{equation}
This is in the form of a pure gauge solution, with gauge parameter
\begin{equation}e^\Lambda = \frac{1}{1-\lambda\Phi}\end{equation}
Note in particular that,
\begin{equation}\frac{d}{d\lambda}\Psi_\lambda = 
d\left(\frac{d\Lambda}{d\lambda}\right)+
\left[\Psi_\lambda,\frac{d\Lambda}{d\lambda}\right]
\end{equation}
so the variation of $\Psi_\lambda$ with respect to $\lambda$ is a gauge 
variation. Since by construction the action is gauge invariant, easy to show 
that,
\begin{equation}\frac{d}{d\lambda}\Tr[\Psi_\lambda d\Psi_\lambda] = 0
\end{equation}
But recalling \eq{SSgauge_sol} we can rewrite this as,
\begin{eqnarray}0&=&\frac{d}{d\lambda}\sum_{m,n=0}^\infty \lambda^{m+n+2}
\langle \psi_m',Q_B\psi_n'\rangle\nonumber\\
&=& \sum_{n=0}^\infty (n+2)\lambda^{n+1}\sum_{k=0}^n
\langle\psi_{n-k}',Q_B\psi_k'\rangle\end{eqnarray} implying 
\eq{diag_sum}. Note that we did not need to assume anything about 
the solution beyond the identities \eq{SSid} and the OSFT axioms.

\section{Arbitrary wedge states}
\label{sec:F}

In this section we investigate a simple generalization of Schnabl's solution
which allows the field $F$ to be an arbitrary wedge state:
\begin{eqnarray}F(K) &=& e^{\gamma K/2}\nonumber\\
&=& \Omega^{\gamma/2}\label{eq:scale_gen}
\end{eqnarray}
where $\gamma\in[0,\infty]$ is a constant defining a one parameter family of
solutions. Note that the second line only follows when $K$ is in the sliver
frame eq.(\ref{eq:Sch_KBc}), which we will assume for simplicity---though the 
analogous generalization exists for any projector frame. More general choices 
of $F$ will be considered in (II). 

The main effect of \eq{scale_gen} is that the strip with insertions defining 
$\psi_n'$ in figure \ref{fig:psi_cor} is ``scaled'' by a factor of $\gamma$.
To see how this effects the solution, consider, generally speaking, two wedge 
states with insertions related by a scaling, $\Phi$ and $\Phi_\gamma$. We 
can represent $\Phi$ as a strip $-\frac{\pi\alpha}{2} + \frac{\pi}{4}<\Re(z)<
\frac{\pi\alpha}{2} - \frac{\pi}{4}$ in the complex plane, with local 
operators,
\begin{equation}\phi_1(z_1),\ \ \phi_2(z_2),\ \ ...\end{equation}
placed inside. The ``scaled'' state $\Phi_\gamma$ is a strip
$\gamma(-\frac{\pi\alpha}{2} + \frac{\pi}{4})<\Re(z)<
\gamma(\frac{\pi\alpha}{2} - \frac{\pi}{4})$ with insertions,
\begin{equation}\phi_1'(z_1'),\ \ \phi_2'(z_2'),\ \ ...\end{equation}
placed inside, where
\begin{equation}z_i'=\gamma z_i\ \ \ \phi_i'(z_i')=s_\gamma\circ\phi_i(z_i)
\ \ \ \ s_\gamma(z) = \gamma z\end{equation}
As explained in ref.\cite{Schnabl}, $\Phi$ can be given an explicit operator
representation,
\begin{equation}|\Phi\rangle = \left(\frac{1}{\alpha}\right)^{\mathcal{L}_0^*}
\left(\frac{1}{\alpha}\right)^{\mathcal{L}_0}\tilde{\phi}_1(z_1)
\tilde{\phi}_2(z_2) ... |\Omega\rangle\end{equation}
where,
\begin{equation}\tilde{\phi}_i(z_i) = f^{-1}\circ\phi_i(z_i),\ \ \ \ \ 
f(z) = \tan^{-1}z \end{equation}
We now consider the contraction between $\Phi$ and a Fock space state
$|\chi\rangle = \tilde{\chi}(0)|\Omega\rangle$:
\begin{equation}\langle \Phi,\chi\rangle = \left\langle \phi_1(z_1)\phi_2(z_2)
... \chi\left(-\frac{\pi\alpha}{2}\right)\right\rangle_{C_{\pi\alpha}}
\end{equation}
Scaling the correlator by a factor of $\gamma$,
\begin{equation}\langle \Phi,\chi\rangle = \left\langle \phi_1'(z_1')
\phi_2'(z_2')... \chi'\left(-\frac{\pi\gamma\alpha}{2}\right)
\right\rangle_{C_{\pi\gamma\alpha}}
\end{equation}
We can see $\Phi_\gamma$ embedded in this, but unfortunately 
the scaled local coordinate has the wrong size to be a Fock space state. 
Nevertheless, this can be regarded as the contraction of $\Phi_\gamma$ with 
another wedge state $\chi_\gamma$ whose strip extends from 
$-\frac{\pi\gamma}{4}<\Re(z)<\frac{\pi\gamma}{4}$ and with a single insertion
$\chi'(0)$:
\begin{eqnarray}\langle \chi_\gamma,\zeta\rangle &=& \left\langle \chi'(0)
\zeta\left(\scriptstyle{-\frac{\pi(\gamma+1)}{4}}\right)
\right\rangle_{C_{\frac{\pi(\gamma+1)}{2}}}\nonumber\\
|\chi_\gamma\rangle &=& \left(\frac{2}{\gamma+1}\right)^{\mathcal{L}_0^*}
\left(\frac{2}{\gamma+1}\right)^{\mathcal{L}_0}\tilde{\chi'}(0)|\Omega\rangle
\label{eq:chi_gamma}
\end{eqnarray}
Thus,
\begin{equation}\langle \Phi,\chi\rangle 
= \langle \Phi_\gamma,\chi_\gamma\rangle\label{eq:phi_phig}\end{equation}
Let us further simplify \eq{chi_gamma}. Note,
\begin{eqnarray}\tilde{\chi'}(0)&=&f^{-1}\circ s_\gamma\circ\chi(0)=
(f^{-1}\circ s_\gamma\circ f)\circ\tilde{\chi}(0)\nonumber\\
&=& \gamma^{\mathcal{L}_0}\tilde{\chi}(0)\gamma^{-\mathcal{L}_0}
\end{eqnarray} The last equality follows from the fact that $\mathcal{L}_0$ 
is the dilatation generator in the sliver frame. Thus \eq{chi_gamma}
becomes,
\begin{equation}|\chi_\gamma\rangle = 
\left(\frac{2}{\gamma+1}\right)^{\mathcal{L}_0^*}
\left(\frac{2\gamma}{\gamma+1}\right)^{\mathcal{L}_0}|\chi\rangle\end{equation}
\eq{phi_phig} then implies,
\begin{equation}\Phi_\gamma = \left(\frac{1+\gamma}{2}\right)^{\mathcal{L}_0}
\left(\frac{1+\gamma}{2\gamma}\right)^{\mathcal{L}_0^*}\Phi
\end{equation}
With a little manipulation of the group algebra 
$[\mathcal{L}_0,\mathcal{L}_0^*]=\mathcal{L}_0+\mathcal{L}_0^*$
\cite{Schnabl,RZ} this can be brought to the form,
\begin{equation}\Phi_\gamma = \exp\left[\frac{1}{2} \ln\gamma(\mathcal{L}_0
-\mathcal{L}_0^*)\right]\Phi\label{eq:mid_rep}\end{equation}
where the operator $\mathcal{L}_0-\mathcal{L}_0^*$ is a midpoint-preserving 
reparameterization generator. It is now simple to see that, 
\begin{equation}\psi_n^{(F=\Omega^{\gamma/2})} = 
\exp\left[\frac{1}{2} \ln\gamma(\mathcal{L}_0
-\mathcal{L}_0^*)\right]\psi_n^{(Schnabl)}\label{eq:psi_scale}\end{equation}
In particular, the factor $1/\gamma$ in \eq{psi_ns} arises from the nontrivial
scaling dimension of the insertions. More generally, we will see in (II) that
$\gamma$ is given by the first order coefficient in the formal Taylor series
expansion,
\begin{equation}F(K) = 1+\frac{\gamma}{2}K +...\end{equation}
At any rate, \eq{psi_scale} implies that the solutions with 
$F=\Omega^{\gamma/2}$ are related by midpoint preserving reparameterizations.
Since these are symmetries of OSFT, it is clear that these solutions are 
physically equivalent.

It may be interesting to consider the limits $\gamma\to 0$ and 
$\gamma\to\infty$. The first limit formally gives an identity based solution, 
while the second gives something resembling the sliver. This suggests a 
connection with vacuum string field theory\cite{VSFT}---indeed in 
ref.\cite{Ghost_Structure} it was argued that vacuum string field theory 
could be derived from OSFT after an infinite midpoint 
reparameterization of the vacuum solution. \Eq{mid_rep} gives an exact 
one parameter family of solutions which realize such a reparameterization
precisely. More work is necessary to understand these limits however, since a 
naive substitution $F=1$ or $F=\Omega^\infty$ results in divergent expressions.

Given the importance of gauge-fixing in the original construction of 
the Schnabl's solution\cite{Schnabl}, it is worth deriving the gauge conditions
satisfied by these solutions. Denoting the solutions by $\Psi_\gamma$, 
Schnabl's solution satisfies,
\begin{equation}\mathcal{B}_0\Psi_1=0\label{eq:Schnabl_gauge}\end{equation}
where $\mathcal{B}_0$ is the $b$ ghost zero mode in the sliver frame. Using
the formula\cite{Schnabl},
\begin{eqnarray}\left(\frac{1}{\alpha}\right)^{-\mathcal{L}_0^*}
\mathcal{B}_0\left(\frac{1}{\alpha}\right)^{\mathcal{L}_0^*} &=&
\frac{1}{\alpha}\mathcal{B}_0 +\left(\frac{1}{\alpha}-1\right)\mathcal{B}_0^*
\end{eqnarray}
we may reparameterize \eq{Schnabl_gauge} to get,
\begin{equation}\left[\frac{\gamma+1}{2}\mathcal{B}_0
+\frac{\gamma-1}{2}\mathcal{B}_0^*\right]\Psi_\gamma =0\end{equation}
Interestingly, in general the gauge condition involves both creation and 
annihilation modes. For the identity-based solution $\gamma=0$ the gauge
fixing operator is a star algebra derivative:
\begin{equation}(\mathcal{B}_0-\mathcal{B}_0^*)\Psi_0=0\end{equation}
Perhaps this indicates a general problem with constructing well-defined 
solutions in $*$-derivative gauges, such as $b_1+b_{-1}=0$. The 
$\gamma\to\infty$ ``vacuum string field theory'' limit gives the condition,
\begin{equation}(\mathcal{B}_0+\mathcal{B}_0^*)\Psi_\infty=0\end{equation}
This gauge also greatly simplifies computation of off-shell amplitudes, as 
was shown in ref.\cite{off-shell}. Interestingly, 
$\mathcal{B}_0+\mathcal{B}_0^*$ happens to annihilate the sliver state, 
giving further evidence for the connection between vacuum string field theory 
and the $\gamma\to\infty$ limit.

\section{Generalizing Conformal Frames}
\label{sec:CFT}

In this section we investigate the second generalization of Schnabl's solution
suggested by the split string formalism: generalizing the choice of vector 
field $v(\z)$, or equivalently, of a projector conformal frame $f(\z)$. For
definiteness we will take the analogue of Schnabl's $F(K)$,
\begin{equation}F(K) = \exp(\half K) \equiv \Xi^{1/2}\end{equation}
The state $\Xi$ depends implicitly on the choice of $f(\xi)$,
and for the sliver frame $\Xi$ is the $SL(2,\mathbb{R})$ vacuum. Beyond the
above formal definition,  $\Xi$ and its powers $\Xi^\alpha$ are generally 
difficult to characterize analytically, for example as of correlators
in the upper half plane.  However, for a certain class of projector 
frames---the {\it special projectors} introduced in ref.\cite{RZ}---the 
$\Xi^\alpha$s can be constructed explicitly.

To make sense of these solutions, the first step should be to calculate the
energy and see whether or not it reproduces the D-brane tension. The main
challenge here is to give a simple definition of these solutions in CFT 
language, so that it is clear how to calculate the inner products necessary
evaluate the energy in terms of CFT correlators. Our strategy will be to 
define a certain conformal frame, the {\it strip frame}, where the 
operator $K_L^v$ becomes a Hamiltonian creating an infinitesimal strip of 
worldsheet. In this formulation the solutions look very similar to Schnabl's 
solution on the cylinder, as in figure \ref{fig:psi_cor}. Moreover, for frames
where $f(\xi)$ diverges only at the midpoint, it is manifest that the 
energy calculation proceeds exactly as for Schnabl's solution. Thus, these 
solutions are simply alternative descriptions of the closed string vacuum. 
Though we will not discuss it explicitly, in fact it turns out that the 
solutions are related by midpoint-preserving reparameterizations\cite{RZO}.

\subsection{The Strip Frame}

\begin{figure}[top]
\begin{center}
{\bf a)}\resizebox{2.5in}{2.1in}{\includegraphics{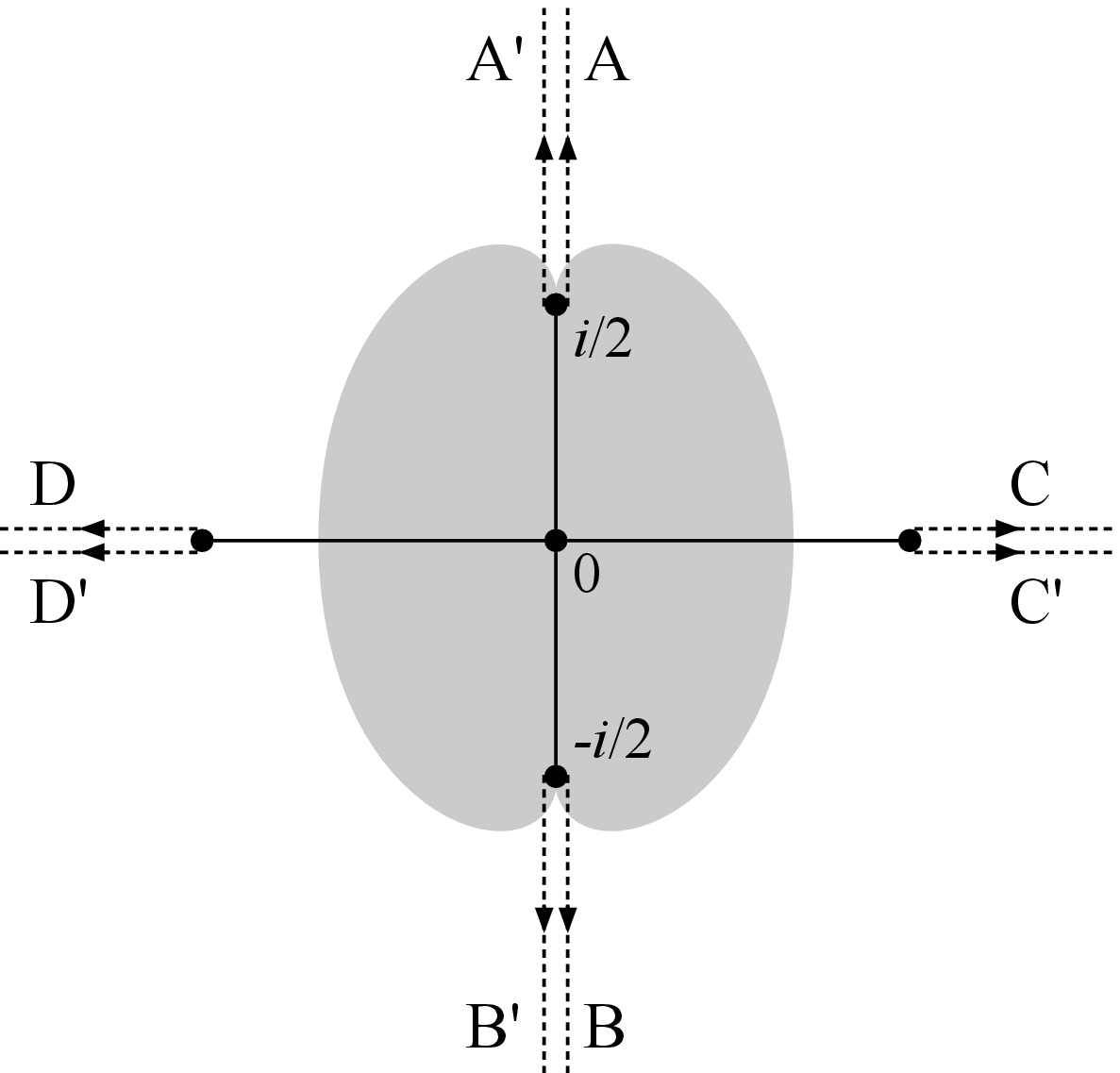}}\ \ \ \ \ \ \ \ \ \ 
{\bf b)}\resizebox{1.6in}{2.1in}{\includegraphics{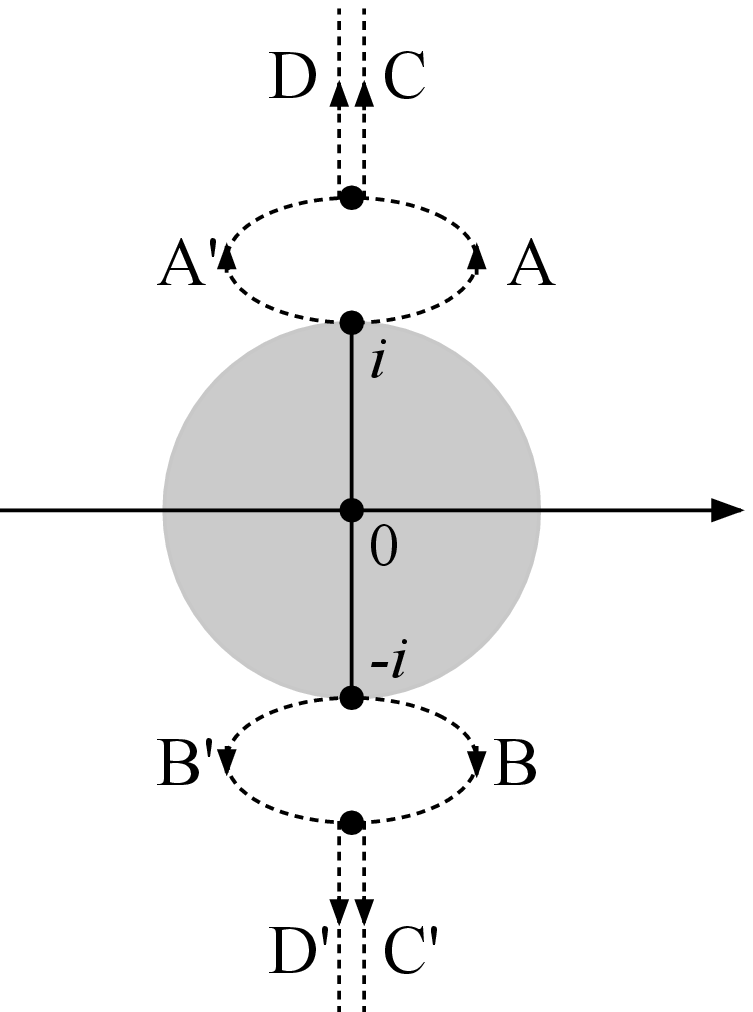}}
\end{center}
\caption{\label{fig:Xi}
{\bf a)} Representation of the surface state $\Xi^\alpha$ in the upper half 
plane, or rather, on the full plane using the doubling trick. The shaded 
region is the local coordinate, and $0$ is the puncture. {\bf b)} 
Picture of $\Xi^\alpha$ in the
$\xi$ coordinate, where the local coordinate is the unit disk. The contours
A,A', etc. which are coincident in the UHP become separated in this 
representation because of branch cuts in $f_\alpha^{-1}$. The surface must
be cut along the dashed lines and glued A$\to$A' etc. to get the appropriate
surface.}
\end{figure}

We begin our analysis by considering an explicit example: the 
butterfly projector, defined by the conformal frame,
\begin{equation}f(\z)= \frac{\z}{\sqrt{1+\z^2}}\label{eq:butterfly}
\end{equation}
This maps the canonical half disk $|\z|<1$ to a region in the upper half plane 
bounded by the branches of the hyperbola $x^2-y^2=\frac{1}{2}$. The vector
field $v(\z)$ and the operators defining the solution become,
\begin{eqnarray}v(\z) &=& \frac{1}{\z}(1+\z^2)^2\nonumber\\
K_L^v &=& (L_{2}+2L_0+L_{-2})_L\ \ \ B_L^v = (b_{2}+2b_0+b_{-2})_L\ \ \ 
c^v(1) = -\frac{1}{4}c(1)
\end{eqnarray}
As discussed in ref.\cite{RZ}, the butterfly is a {\it special projector}, 
meaning the energy-momentum zero modes $\mathcal{L}_0,\mathcal{L}_0^*$ in 
this frame satisfy the algebra,
\begin{equation}[\mathcal{L}_0,\mathcal{L}_0^*] = s(\mathcal{L}_0
+\mathcal{L}_0^*)\end{equation}
where for the butterfly $s=2$. This extra structure gives us a little more 
information about how to construct the analogue of wedge states for the 
butterfly, the $\Xi^\alpha$s. Explicitly, they may be represented as surface 
states on the upper half plane, with a local coordinate\cite{RZ}
\begin{equation}\mu = f_\alpha(\z) = 
\frac{\z\sqrt{1+2\alpha(1+\z^2)}}{1-\z^2 +2\alpha(1+\z^2)}\end{equation}
Also useful is the inverse map, 
\begin{equation}\z = f^{-1}_\alpha(\mu) = 
\sqrt{\frac{2\alpha+1}{2}}\left(\frac{1-2(2\alpha-1)\mu^2
-\sqrt{4\mu^2-1}}{(2\alpha-1)^2\mu^2-2\alpha}\right)^{1/2}\label{eq:fa_inv}
\end{equation}
A picture of a typical $\Xi^\alpha$ surface state is shown in figure 
\ref{fig:Xi}, both on the upper half plane and in the $\z$-presentation 
(where the local coordinate is the unit disk). In the $\z$-presentation there
are various cuts in the surface, one surrounding a pseudo-circular region just
above the local coordinate and another further above on the imaginary axis.
These cuts must be glued together as indicated before the 
$\Xi^\alpha$ surface is formed. Neither coordinate system, known from 
ref.\cite{RZ}, is well suited for calculating star products or vertices. 

\begin{figure}[top]

\begin{center}
{\bf a)}
\resizebox{2.1in}{1.6in}{\includegraphics{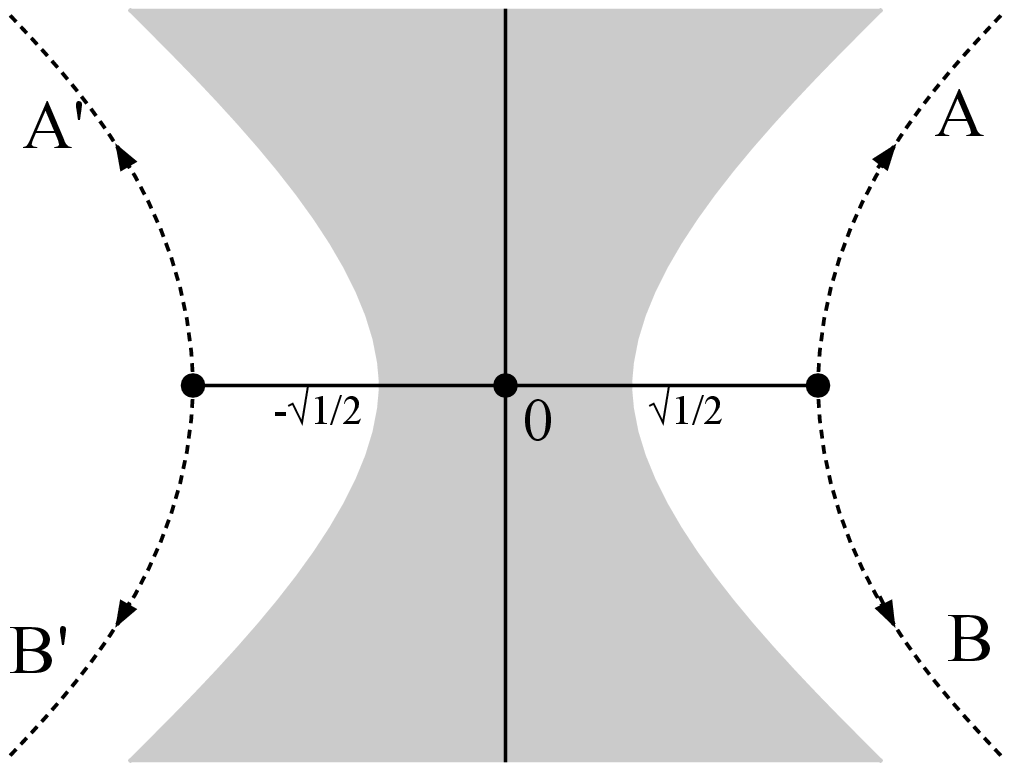}}\ \ \ \ \ \ \ \ \ \ 
{\bf b)}\resizebox{1.7in}{1.5in}{\includegraphics{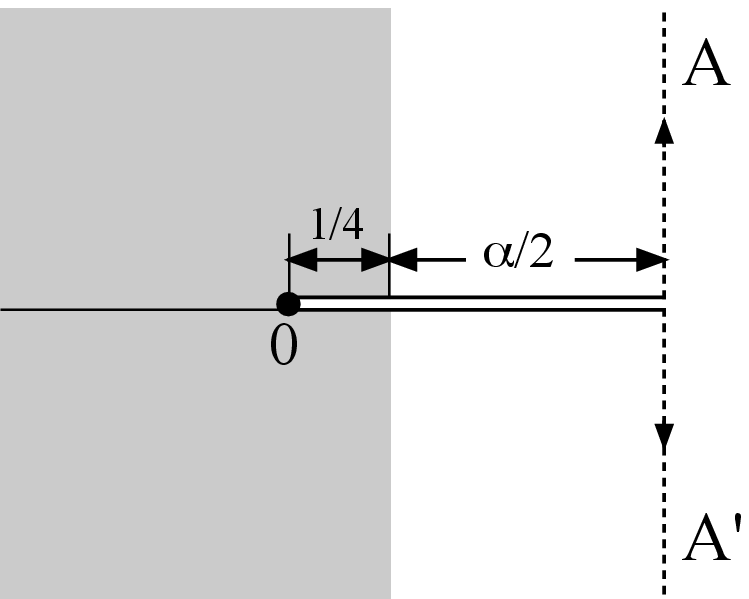}}\ \ \ \ \ \ \ \ \ \
{\bf c)}\resizebox{1.0in}{2.0in}{\includegraphics{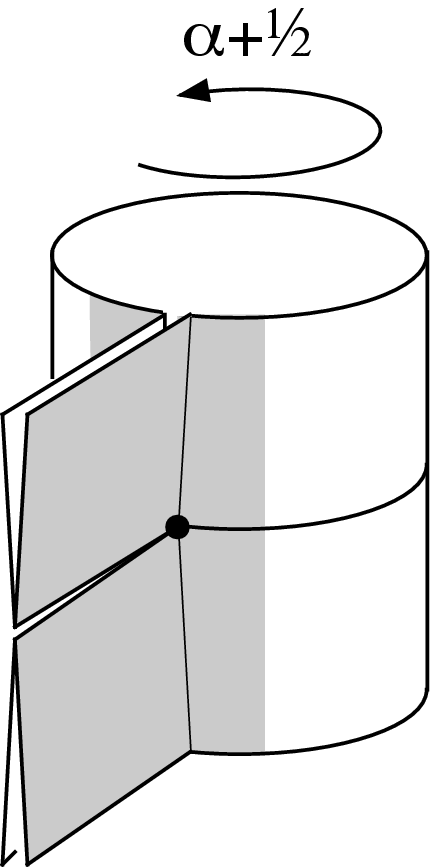}}
\end{center}
\caption{\label{fig:Xi_proj}
{\bf a)} Representation of $\Xi^\alpha$ in the butterfly frame. The dashed
hyperbolic branches A,A',B,B' are the images of the corresponding dashed 
pseudo-circles above and below the local coordinate in fig.\ref{fig:Xi}b. 
The surface must be cut and glued along these branches as indicated.
{\bf b)} The $\Xi^\alpha$s in the strip frame, $H=\half f^2$. 
For clarity we have
only shown the image of the upper half plane. The semi-infinite grey shaded 
region $\Re(z)<\frac{1}{4}$ is the local coordinate, i.e. the image of the 
upper half unit semi-circle in the $\xi$-presentation. The dashed lines A,A' 
are the images of the hyperbolic branches in a) and should be glued together. 
The boundary of the open string in this picture is above and below the positive
real axis, and if we use the doubling trick the image of lower half plane is 
on another Riemann sheet glued along the positive real axis. 
{\bf c)} After gluing 
A,A' and including the lower half plane, we obtain a picture of a cylinder
of circumference $\alpha+\frac{1}{2}$ together with two semi-infinite sheets
folded in half and glued parallel to the axis of the cylinder and meeting at
the puncture.}
\end{figure}

The description of $\Xi^\alpha$ simplifies somewhat in the butterfly frame, 
as shown in figure \ref{fig:Xi_proj}a. After transforming with 
\eq{butterfly}, the pseudo-circular region gets mapped to the exterior
of a hyperbola $x^2-y^2=\half +\alpha$, lying just outside of the butterfly
local coordinate. The branches of the outer hyperbola must be sewn together to
form a kind of warped cylinder. Though an improvement over figure \ref{fig:Xi},
these hyperbolic contours are awkward for gluing. There is a related 
issue here: 
Unlike the case for the sliver, the operator $K_L^v$ in the butterfly frame 
is not simply a ``Hamiltonian'' which creates an infinitesimal strip. Rather,
\begin{equation}f\circ K_L^v = 
\int_{i\infty}^{-i\infty}\frac{dz}{2\pi i}\frac{1}{z}T(z)\end{equation}
This suggests that we should look for another conformal frame where $K_L^v$
simply becomes $\int \frac{dz}{2\pi i}T(z)$. We will call this the 
``strip frame'' $y=H(\z)$. We want,
\begin{equation}H^{-1}\circ \int \frac{dy}{2\pi i}T(y) = 
\int_L \frac{d\z}{2\pi i}\frac{1}{H'(\z)}T(\z) = K_L^v \end{equation}
In other words,
\begin{equation}\frac{d}{d\z}H(\z) = \frac{1}{v(\z)}\label{eq:strip_frame}
\end{equation}
This equation is easily solved in the situation at hand. The result is,
\begin{equation}H(\z) = \frac{1}{2}\frac{\z^2}{1+\z^2} = \frac{1}{2}f(\z)^2
\label{eq:butt_strip}\end{equation}
That is, we square $f(\z)$ and divide by 2. The resulting picture of 
$\Xi^\alpha$ is very simple, as shown in figure \ref{fig:Xi_proj}b. 
Concentrate for the moment on the image of the upper half plane. The 
hyperbola-shaped local coordinate gets mapped to the entire region 
$\Re(z)<\frac{1}{4}$, whose positive and negative boundaries are now
straight vertical lines above and below the real axis. The oddly shaped
regions between the the cut and local coordinate are
mapped to rectangular strips of width $\alpha/2$ above and below the real 
axis. Star multiplication of $\Xi^\alpha$s is clear: it simply changes the 
width of these strips\footnote{If 
we use the identification to bring the entire strip to the positive boundary 
of the local coordinate, we 
have chosen our conventions so that $\Xi^\alpha$ always corresponds to a 
region of width $\alpha$ in the strip frame. For wedge states, the strip 
frame is related to the sliver frame by a factor of $2/\pi$, so this is 
consistent with \eq{wedge}.}. 
Gluing the strips together and sewing the image of the lower half plane onto 
positive real axis, we get the surface pictured in \ref{fig:Xi_proj}c. This 
is almost a cylinder of circumference $\alpha+\frac{1}{2}$, except there 
are two semi-infinite planes inside the local coordinate folded in half and 
glued parallel to the axis, joining at the puncture.

One minor subtlety with this coordinate system is the presence of a 
curvature singularity at the puncture. In particular, inserting an off-shell 
vertex operator with $H(\xi)$ produces a divergent factor from 
the conformal transformation. Of course, this singularity
is fictitious since in the end we must map back to the upper half plane to 
calculate the correlator anyway. But to be proper, we should define the surface
state in the strip frame as a limit,
\begin{equation}\langle \Xi^\alpha,\chi\rangle = 
\lim_{\z\to 0}\langle H\circ\chi(\z)\rangle_{X_{\alpha+\frac{1}{2}}}
\end{equation} 
where we denote,
$$X_{\alpha+\frac{1}{2}}=H\circ f_\alpha^{-1}\circ\mathrm{UHP},$$ 
the surface pictured in figure \ref{fig:Xi_proj}b,c.

\begin{figure}[top]
\begin{center}
\resizebox{2.8in}{1.5in}{\includegraphics{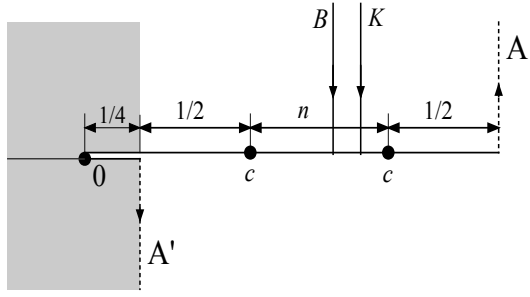}}
\end{center}
\caption{\label{fig:psin_Xi}
The $\psi_n'$ building blocks of the butterfly generalization of 
Schnabl's solution. The entire strip has been glued to the positive 
boundary of the local coordinate, and only the image of the upper half
plane is shown for clarity.}
\end{figure}

It is a simple exercise to express the $\psi_n'$s in the strip frame,
as shown in figure \ref{fig:psin_Xi}:
\begin{equation}\langle \psi_n',\chi\rangle = \lim_{\z\to 0}\left\langle 
c\left(n+\frac{3}{4}\right)KB c\left(\frac{3}{4}\right)H\circ\chi(\z)
\right\rangle_{X_{n+\frac{3}{2}}}\label{eq:psin_Xi}
\end{equation}
with the $K,B$ contour insertions as in \eq{KB_cont}. Note that care
must be taken to insert the operators on the proper branch. In 
\eq{psin_Xi} and figure \ref{fig:psin_Xi} we attached the entire strip to 
the positive boundary of the local coordinate, so no confusion arises. 
However, if we were to attach half the strip to the negative boundary, as 
in figure \ref{fig:Xi_proj}b, both $c$s would be located at $\frac{3}{4}$; of 
course, this does not give zero because the insertions are on different 
branches. In this coordinate system, it is easy to prove the equations of 
motion by following step-by-step the CFT derivation given in ref.\cite{Okawa}.
Furthermore, it is clear that the inner products 
$\langle \psi_m',Q_B\psi_n'\rangle$ will be the 
same as for Schnabl's solution, since after we cut away the local coordinates 
and glue the strips together we are left with the same correlators on the 
cylinder. Therefore, the the butterfly solution reproduces the expected 
D-brane tension.

Let us discuss how these results generalize to other projector frames. It 
turns out that $H(\xi)$ satisfies an interesting reality condition which plays
a particularly important role in this respect. Note that the reality condition
for $v(\xi)$ \eq{v_real} implies:
\begin{eqnarray}\overline{\frac{d}{d\z}H(\z)} = \frac{1}{\,\overline{v(\z)}\,}
=\frac{1}{\bar{\z}^2 v\left(\frac{1}{\bar{\z}}\right)} = 
-\frac{d}{d\bar{\z}}H\left(\frac{1}{\bar{\z}}\right)
\end{eqnarray}
Integrating this yields,
\begin{equation}\overline{H(\z)} + H\left(\frac{1}{\bar{\z}}\right) = 
2 A
\end{equation}
where $A$ is a real integration constant. When $\z = e^{i\theta}$ is on the 
unit circle, this implies:
\begin{equation}\mathrm{Re}H(e^{i\theta}) = A\end{equation}
That is, the boundary of the local coordinate in the strip frame is always a 
straight, vertical line intersecting the real axis at $A$. Furthermore, 
because $K_L^v$ creates pieces of worldsheet of constant 
width parallel to the imaginary axis, the states $\Xi^\alpha$ always 
correspond to strips of width $\alpha$ glued to the positive boundary of the
local coordinate, regardless of the choice of projector. Therefore it is 
clear that the split string solution \eq{SSsol} reproduces the correct 
D-brane tension regardless of the choice of vector field 
$v(\xi)$\footnote{modulo some singularities which we consider in the next 
subsection}. However for completely general $v(\xi)$ the local coordinate can 
be an extremely complicated Riemann surface, subject only to the constraint 
that its boundary is a vertical line in the complex plane. Thus it is
difficult to perform explicit calculations with the solution, for example
in level truncation.

For special projectors, however, the story drastically simplifies. Special 
projectors satisfy some additional constraints which yield a simple 
algebraic relationship between $H(\z)$ and the coordinate frame defining the
projector. As explained in ref.\cite{RZ}, all special conformal frames 
$f(\z)$ satisfy the constraint,
\begin{equation}\left[-f\left(-\frac{1}{\z}\right)\right]^s + C_1 f(\z)^s
=C_2\end{equation}
where $C_1$ and $C_2$ are constants over the unit circle, except possibly 
at points where $f(\z)$ becomes singular. We can use this to simplify $v(\z)$,
\begin{eqnarray}v(\z) &=& \frac{f(\z)}{f'(\z)}-
\frac{f(-1/\z)}{\frac{d}{d\z}f(-1/\z)} \nonumber\\
&=& \frac{f}{f'}- \frac{(C_2-C_1f^s)^{1/s}}{\frac{d}{d\z}(C_2-C_1f^s)^{1/s}}
\nonumber\\
&=& \frac{C_2}{C_1f^{s-1}f'} = \frac{1}{H'}\end{eqnarray}
This equation integrates easily to give
\begin{equation}H(\z)= \frac{C_1}{C_2s}f(\z)^s,\label{eq:strip_special}
\end{equation}
a generalization of \eq{butt_strip}.

\begin{figure}[top]
\begin{center}
\resizebox{5in}{2.0in}{\includegraphics{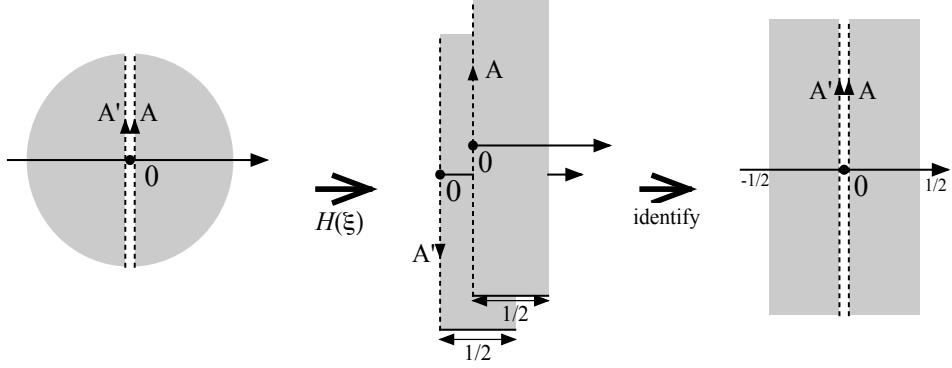}}
\end{center}
\caption{\label{fig:sliver}
Mapping the local coordinate using the sliver's $H(\xi)$. The two halves of the
unit disk get mapped to two copies of the strip $\Re(z)\in[0,\frac{1}{2}]$. The
contours A,A' which are coincident on the disk must be glued so that the
arrows point in the same direction. The result is the familiar picture of the
sliver frame as a strip $\Re(z)\in[-\frac{1}{2},\frac{1}{2}]$.}
\end{figure}

Let us see how \eq{strip_special} determines the structure of the local 
coordinate in the strip frame. Using constraints on $C_1,C_2$ 
derived in ref.\cite{RZ} we can show,
\begin{equation}\mathrm{Re}H(e^{i\sigma}) = \frac{1}{2s}\end{equation}
So the boundary of the local coordinate lies on a straight vertical line
intersecting the real axis at $\frac{1}{2s}$. In fact, because the local 
coordinate reaches infinity at the midpoint, as we move around the unit
circle from $\sigma=0$ to $2\pi$ we will pass over this vertical line
at least\footnote{If $f(\xi)$ has singularities elsewhere than at the 
midpoint, there may be many copies of this line---see next section.} twice.
This means that the local coordinate patch lies on two or more Riemann 
sheets with boundary on this line---we have seen this phenomenon already for 
the butterfly. Next, note that the ``constants'' $C_1,C_2$ can be discontinuous
on the unit circle where $f(\xi)$ has singularities. We may extend the 
definition of $C_1,C_2$ for $|\xi|<1$ by defining them to be piecewise constant
on pie shaped regions inside the unit circle. These discontinuities then
imply that the local coordinate patch in the strip frame generally has cuts 
extending from the origin (the puncture) to infinity, with implicit 
identifications. The simplest example of this is the sliver 
$f(\z) = \tan^{-1}\z$, where,
\begin{equation}C_1 = 1\ \ \ C_2 = \frac{\pi}{2}\epsilon(\z)\end{equation}
and $\epsilon(\z)$ is the step function, $=1$ for $\mathrm{Re}(\z)>0$ and
$-1$ for $\mathrm{Re}(\z)<0$. Plugging in to \eq{strip_special} then gives,
\begin{equation}H(\z) = \frac{2}{\pi}\epsilon(\z)\tan^{-1}\z\end{equation}
This maps the unit disk to two copies of the strip 
$\mathrm{Re}(z)\in[0,\frac{1}{2}]$, as shown in figure 
\ref{fig:sliver}. Note that the local coordinate lies on two Riemann 
sheets, as anticipated above. On the negative boundaries of these strips are 
cuts extending from the origin to infinity; these cuts are images of of the 
contours on either side of the imaginary axis on the disk. Sewing these cuts 
together, we recover the usual picture of the sliver local coordinate 
(up to a scaling) as a single strip, 
$\mathrm{Re}(z)\in[-\frac{1}{2},\frac{1}{2}]$.

\subsection{Multi-Winged Butterflies}

\begin{figure}[top]
\begin{center}
\resizebox{2.8in}{1.5in}{\includegraphics{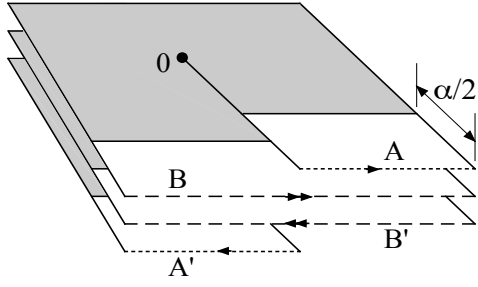}}
\end{center}
\caption{\label{fig:Xi_multi}
Riemann sheets of the state $\Xi^\alpha$ in the strip frame for 
the $m=3$ butterfly. Shaded in grey is the local coordinate, and in white are
the attached strips. The topmost and bottom most sheets A,A' must be 
identified so that the arrows point in the same direction.
Likewise, the middle two sheets B,B' must be identified so that the double 
arrows point in the same direction. This picture only shows the image of the 
upper half plane; there is a similar picture for the lower half plane which 
must be glued to the line segments representing the boundary of the open 
string. This gives a total of 6 complete branches.}
\end{figure}

Let us give another class of examples which seem fundamentally different from
the sliver and butterfly. We will call these ``multi-winged'' butterflies, and
they are defined by the conformal frames\cite{Projectors,RZ}:
\begin{equation}f_{(m)}(\z) = 
\frac{\z}{(1+(-1)^{m+1}\z^{2m})^{1/2m}}\end{equation}
for integer $m\geq 1$ and have $s=2m$; $m=1$ is the butterfly. Calculating 
$v(\z)$ and the strip frame,
\begin{eqnarray}v_{(m)}(\z)&=&\z\left[2+(-1)^{m+1}(\z^{2m}+\z^{-2m})\right]
\nonumber\\
H_{(m)}(\z) &=& \frac{1}{2m}\frac{\z^{2m}}{\z^{2m}+(-1)^{m+1}} = 
\frac{(-1)^{m+1}}{2m}[f^m(\z)]^{2m}
\end{eqnarray}
consistent with the general structure \eq{strip_special}. We can also
work out the image of the unit circle in the strip frame,
\begin{eqnarray}H_{(m)}(e^{i\theta}) &=& \frac{1+i\tan m\theta}{4m}\ \ \ m\ 
\mathrm{odd}\nonumber\\ &=& \frac{1-i\cot m\theta}{4m}\ \ \ m\ \mathrm{even}
\end{eqnarray} The image curves have constant real part $1/4m$ and so are 
straight lines, as expected. The interesting thing about these frames is that
when $\theta$ increases over the left half of the string, the coordinate 
passes through infinity and over the same vertical line 
many times. This is a consequence of the fact that the frames are
singular at multiple points on the unit circle,
\begin{equation}\theta_k = \frac{k}{m}\frac{\pi}{2}\end{equation}
where $k$ is odd when $m$ is odd and $k$ is even when $m$ is even. The 
appearance of many copies of the same vertical line indicates that we should 
think of the local coordinate in terms of multiple Riemann sheets. The 
states $\Xi^\alpha$ are obtained by gluing strips of length $\alpha/2$ to the 
branches of the boundary of the local coordinate, as shown in figure 
\ref{fig:Xi_multi}. The situation is analogous to the butterfly, only now
we have ``many wings.''

Let us now try to make solutions in these frames. Note that when $m$ is even,
an odd thing happens: the boundary of the open string is mapped to $-i\infty$.
When constructing $\psi_n$s we need to place $c$ ghosts on the boundary, so 
this requires ghost insertions in the strip frame at $-i\infty$.
We can see a related problem in the split string formalism. When $m$ is even 
$v(1)=0$, so the corresponding $c$ field is divergent. Perhaps these solutions
can be regulated by temporarily relaxing reality and taking the ghosts off 
the real axis.

When $m$ is odd, however, the solutions seem better, at least in the
Fock space. Take for example the $\psi_n'$s when $m=3$. Moving everything to 
the positive boundary of the local coordinate, we obtain three strips of 
width $n+1$. One strip is bisected by 
the open string boundary; we place $c$ insertions and $K,B$ contours inside 
exactly as in figure \ref{fig:psin_Xi}. The other two strips contain the 
images of the unit circle for $\theta\in[\frac{\pi}{6},\frac{\pi}{2}]$ and 
$\theta\in[-\frac{\pi}{2},-\frac{\pi}{6}]$, respectively; these strips 
contain the remaining pieces of the $K,B$ contours. Calculation of the energy,
however, brings a puzzling phenomenon: upon removing the local coordinate,
the inner product $\langle\psi_n',Q_B\psi_n'\rangle$ involves a correlator on
three disconnected cylinders! We were not able to follow the analysis far 
enough to come to a definite conclusion about these solutions, but there 
could be an interesting phenomenon here.

\section{Conclusion}

In this paper we have explored some generalizations of Schnabl's solution 
suggested by the split string formalism. All of the solutions we have analyzed
turn out to be related by midpoint preserving reparameterizations\cite{RZO};
indeed, with the help of the strip frame \eq{strip_frame} it is manifest that
these solutions are physically equivalent. Less obvious is what happens 
when we let $F$ be a completely general element of the Abelian algebra of 
wedge states; we will consider this question in the companion paper (II).

Originally our hope was that the split string solution eq.(\ref{eq:SSsol}) 
would be general enough to accommodate multiple brane vacua, though this 
possibility seems not to have been realized. It is possible that solutions 
based on multi-branched projector frames could bring surprises, or perhaps
a completely new understanding of the $\psi_N$ piece is necessary. Another 
possible approach would be to explore the vacuum string field theory limit, 
where multiple brane solutions are better understood.

The author would like to thank D. Gross and A. Sen for conversations. This
work was supported by the Department of Atomic Energy, Government of India.

\end{document}